\documentclass[twocolumn,conference]{IEEEtran}

\usepackage{graphicx}
\usepackage{amsmath}
\usepackage{amssymb}
\usepackage{algorithmic}
\usepackage{algorithm}
\usepackage{multirow}

\begin{document}

\title{Surface Reconstruction from Scattered Point via RBF Interpolation on GPU}

\author{
\IEEEauthorblockN{Salvatore Cuomo\IEEEauthorrefmark{1}, Ardelio
Galletti\IEEEauthorrefmark{2}, Giulio Giunta\IEEEauthorrefmark{2},
Alfredo Starace\IEEEauthorrefmark{2}}
\IEEEauthorblockA{\IEEEauthorrefmark{1}Department of Mathematics and
Applications ``R. Caccioppoli''
University of Naples Federico II\\
c/o Universitario M.S. Angelo 80126 Naples Italy
\\
email:salvatore.cuomo@unina.it}
\IEEEauthorblockA{\IEEEauthorrefmark{2}Department of Applied Science.
University of Naples ``Parthenope''.\\Centro Direzionale, Isola C4 80143 Naples Italy\\
emails:\{ardelio.galletti,giulio.giunta\}@uniparthenope.it,
alfredo.starace@gmail.com} }

\maketitle

\begin{abstract}
In this paper we describe a parallel implicit method based on radial
basis functions (RBF) for surface reconstruction. The applicability
of RBF methods is hindered by its computational demand, that
requires the solution of linear systems of size equal to the number
of data points. Our reconstruction implementation relies on parallel
scientific libraries and is supported for massively multi-core
architectures, namely Graphic Processor Units (GPUs). The
performance of the proposed method in terms of accuracy of the
reconstruction and computing time shows that the RBF interpolant can
be very effective for such problem.

\end{abstract}

\section{Introduction}\label{sec:Introduction}

Many applications in engineering and science need to build accurate digital models of real-world objects defined in terms
of \emph{point cloud data}, \emph{i.e.} a set of scattered points in 3D. Typical examples include the digitalization of
manufactured parts for quality control, statues and artifacts in archeology and arts \cite{Levoy2000}, human bodies for
movies or video games, organs and anatomical parts for medical diagnostic \cite{carr1997surface} and elevation models
for simulations and modeling \cite{Pouderoux2004}. Using modern 3D scanners, it is possible to acquire point clouds
containing millions of points sampled from an object. The process of building a geometric model from such point clouds
is usually referred to as \emph{surface reconstruction}.

There are several approaches to reconstruct surfaces from 3D
scattered datasets. Generally,  the methods of surface
reconstruction fall into two categories \cite{schall2005surface}:
Delaunay-based methods and volumetric and implicit based methods.
Delaunay triangulation is usually utilized to find the possible
neighbors for each point in all directions from all samples.
Implicit surface modeling instead is most popular for describing
complex shapes and complex editing operations. Among them, level set
methods \cite{zhao2001fast}, moving least square methods
\cite{klein2004point}, variational implicit surfaces
\cite{turk1999variational} and adaptively sampled distance field
\cite{frisken2000adaptively} are recent developments in this field.
In this paper we  focus on the implicit surface method based on
radial basis functions (RBFs). In the 1980's,
Franke \cite{franke1982scattered} used radial basis functions to
interpolate scattered point cloud firstly and proved the accuracy
and stability of the interpolation based on RBFs. Using this
technique, an implicit surface is constructed by calculating the
weights of a set of radial basis functions whose linear combination
interpolates the given data points.

The RBF applicability is hindered by its computational demand, since
these methods require the solution of a linear system of size equal
to the number of data points and current 3D data scanners allow
acquisition of tens of millions points of an object surface.

High Performance Computing (HPC) is a natural solution to provide
the computational power required in such situations. In this paper
we propose a method designed for a massively multi-core
architecture, namely Graphics Processing Units (GPUs)
\cite{sussmuth2010surface}. Recently, GPUs have been effectively
used to accelerate the performance of applications in several
scientific areas such as computational fluid dynamics, molecular
dynamics, climate modeling \cite{SCuomo1}. For our knowledge, the
most efficient parallel algorithm for RBF interpolation is
``PetRBF'' \cite{yokota2010petrbf}. PetRBF is a parallel algorithm
for RBF interpolation that exhibits $\mathcal{O}(N)$ complexity,
requires $\mathcal{O}(N)$ storage, and scales excellently up to a
thousand processes. Our main contribute is a deep re-engineering of
PetRBF which constitutes a generalization for scattered 3D data and
an extension for GPU acceleration on heterogeneous clusters. We also
focus on the suitable choice of the algorithm  parameters and
present an optimal strategy for synthetic, real or incomplete
datasets.

In section II we deal mathematical Preliminaries, in section III we
describe the related works and the implementation strategies. The
section IV we report the numerical experiments and finally we draw
conclusions in section V.

\section{Preliminaries}
In this section we recall the reconstruction based on implicit
surface method and define the related  RBF interpolation problem.

\subsection{Implicit Surface Reconstruction}
Given a point cloud $$\mathcal{X}
:=\{(x_i,y_i,z_i)\in\mathbb{R}^3,i=1,\dots,N \}$$ coming from an
unknown surface $\mathcal{M}$, \emph{i.e.} $\mathcal{X} \subset
\mathcal{M}$, the goal is to find another surface $\mathcal{M}^*$
which is a reconstruction of $\mathcal{M}$. In the implicit surface
approach $\mathcal{M}$ is defined as the surface of all points
$(x,y,z) \in \mathbb{R}^3$ that satisfy the implicit equation
\begin{equation}
\label{eq:impl}
f(x,y,z)=0
\end{equation}
for an unknown function $f$. A way to approximate $f$ is to impose
the interpolation conditions (\ref{eq:impl}) on the point cloud
$\mathcal{X}$. However, the use of  those interpolation conditions
only leads to the trivial solution given by the identically zero
function, whose zero surface is $ \mathbb{R}^3$. Therefore, the key
for finding an approximation of the function $f$ is to use
additional significant interpolation conditions, \emph{i.e.} on
correspondence of off-surface points (where $f \not=0$). This
involves a nontrivial interpolant $\mathcal{P}_f$, whose zero
surface contains a meaningful surface $\mathcal{M}^*$. This approach
leads to a surface reconstruction method which consists of three
main steps:
\begin{enumerate}
\item generation of off-surface points;
\item interpolant model identification on the extended dataset;
\item computation of the interpolation zero iso-surface.
\end{enumerate}

\subsubsection{Generation of off-surface points}
\begin{figure}[h!]
\center
 \includegraphics[width=.4\textwidth]{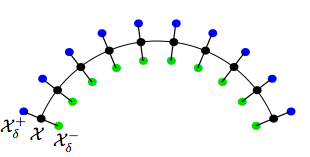}\\
  \caption{Extended interpolation data set. In black points from $\mathcal{X}$, in blue points from $\mathcal{X}_\delta^+$ and in green points from $\mathcal{X}_\delta^-$}\label{img:extdset}
\end{figure}

\ \\
A common practice, as suggested in \cite{turk2001implicit}, is to
use the set of surface normals $\mathbf{n}_i=(n_i^x,n_i^y,n_i^z)$ to
the surface $\mathcal{M}$ at points $\mathbf{x}_i=(x_i,y_i,z_i)$. If
these normals  are not explicitly
known, there are techniques and tools\footnote{package ply.tar.gz provided by Greg Turk, available at \\
http://www.cc.gatech.edu/projects/large\_models/ply.html} that to
allow to estimate them. Given the oriented surface normals
($\mathbf{n}_i$ and $-\mathbf{n}_i$), we generate the extra
off-surface points by marching a small distance $\delta$ along the
normals. So, we obtain for each cloud data point
$\mathbf{x}_i=(x_i,y_i,z_i)$ two additional off-surface points. The
first lies ``outside'' the surface $\mathcal{M}$ and is given by
\begin{equation*}
\begin{split}
(x_{N+i},y_{N+i},z_{N+i})=\mathbf{x}_i+\delta\mathbf{n}_i=\ \ \ \ \ \ \ \\
=(x_i+\delta n_i^x,y_i+\delta n_i^y,z_i+\delta n_i^z);
\end{split}
\end{equation*}
the other point lies ``inside'' and is given by
\begin{equation*}
\begin{split}
(x_{2N+i},y_{2N+i},z_{2N+i})=\mathbf{x}_i-\delta\mathbf{n}_i= \ \ \ \ \ \\
=(x_i-\delta n_i^x,y_i-\delta n_i^y,z_i-\delta n_i^z).
\end{split}
\end{equation*}
The union of the sets
$\mathcal{X}_\delta^+=\{\mathbf{x}_{N+1},\dots,\mathbf{x}_{2N}\}$,
$\mathcal{X}_\delta^-=\{\mathbf{x}_{2N+1},\dots,\mathbf{x}_{3N}\}$
and $\mathcal{X}$ gives the overall set of points on which the
interpolation conditions are assigned (see Fig. \ref{img:extdset}).
The set $\mathcal{X}_\delta^+$ implicitly defines a surface
$\mathcal{M}_\delta^+$ which passes through its points. Analogously
$\mathcal{X}_\delta^-$ defines the surface $\mathcal{M}_\delta^-$.
Those two surfaces can be considered respectively external and
internal to $\mathcal{M}$. The value of $\delta$ represents a small
step size whose specific magnitude may be rather critical for a good
surface reconstruction \cite{carr2001reconstruction}. In particular,
if $\delta$ is chosen too large, this
 results in self intersecting $\mathcal{M}_\delta^+$ or $\mathcal{M}_\delta^-$ auxiliary surfaces. In our implementation we
 fix $\delta$ to $1\%$ of the bounding box of the data as suggested in \cite{wendland2005scattered}.

\subsubsection{Interpolant model identification on extended dataset}
\ \\
This step consists in determining a function $\mathcal{P}_f$ whose
zero contour interpolates the given point cloud data $\mathcal{X}$
and whose iso-surface $\mathcal{P}_f=1$ and $\mathcal{P}_f=-1$
interpolate $\mathcal{X}_\delta^+$ and $\mathcal{X}_\delta^-$,
respectively, \emph{i.e.}
$$
\mathcal{P}_f(x_i)=\left\{
\begin{array}{rl}
0&\ \ \ \ i=1,\dots,N\\
1&\ \ \ \ i=N+1,\dots,2N\\
-1&\ \ \ \ i=2N+1,\dots,3N
\end{array}
\right.
$$
The values of $\pm 1$ for the auxiliary data are assigned in an
arbitrary way. Such  choice does not affect the quality of the
results. In this discussion we are interested to the iso-surface
zeros of $\mathcal{P}_f$.

\subsubsection{Computation of the interpolation zero iso-surface}
\ \\
In order to evaluate the $\mathcal{P}_f$ zero iso-surface and
visualize it, we use a simple strategy which consists on evaluating
the interpolant $\mathcal{P}_f$ on a dense grid of a bounding box.
This approach leads to some undesired artifacts, since in the
bounding box there are points that do not belong to $\mathcal{M}^*$.
A possible way to overcome this drawback and display only
$\mathcal{M}^*$ consists in evaluating the interpolant only in a
small \emph{surrounding} volume of the surface $\mathcal{M}$ to
reconstruct. This set is denoted as
$\mathcal{M}_{ext}^\varepsilon=\{\mathbf{x}\in\mathbb{R}^3:\
d(\mathbf{x},\mathcal{M})\leq\varepsilon\}$, where
$d(\mathbf{x},\mathcal{M}) =
\displaystyle\inf_{\mathbf{y}\in\mathcal{M}}
\Vert\mathbf{y}-\mathbf{x}\Vert$. For a small enough value of
$\varepsilon$ it holds
$$
\mathcal{M}^*\approx\mathcal{M}_{ext}^\varepsilon\cap\mathcal{S}_0.
$$

\noindent where $ \mathcal{S}_0$ is the zero iso-surface of $\mathcal{P}_f$.

\subsection{RBF interpolation}
Given a set of $N$ distinct points $(x_j,\ y_j)$, $j=1,\dots,N$, where $x_j \in \mathbb{R}^s$ and $y_j \in \mathbb{R}$,
the scattered data interpolation problem is to find an interpolant function $\mathcal{P}_f$ such that:
\begin{equation}
\label{eq:interp_cond}
\mathcal{P}_f(x_j) = y_j,\ j=1,\dots,N.
\end{equation}
In the univariate setting, the interpolant $\mathcal{P}_f$ is
usually chosen in suitable spaces of functions. A common approach
assumes that the function $\mathcal{P}_f$ is a linear combination of
certain basis function $B_j$, \emph{i.e.}
\begin{equation}
\label{eq:base_exp}
\mathcal{P}_f(x)=\sum_{j=1}^N c_jB_j(x).
\end{equation}
In the multivariate setting ($x_j\in\mathbb{R}^s,\ s>1 $), however, the problem is much more complex. As stated by the \emph{Mairhuber-Curtis} theorem \cite{curtis1959n,mairhuber1956},
 in order to have a well-posed multivariate scattered data interpolation problem  it is not possible to fix in advance the
basis $\{B_1,\dots,B_N\}$. Instead the basis must depend on the data location.

In order to obtain data dependent approximation space, as suggested
by the Mairhuber-Curtis theorem, the RBF interpolation uses radial
functions: $$B_j\equiv\Phi_j=\varphi(\Vert x-x_j\Vert).$$
The points $x_j$ to which the basic function $\varphi$ is shifted
are usually referred as \emph{centers}. While there may be
circumstances that suggest to choose these centers different from
the data sites one generally picks the centers to coincide with the
data sites.

The interpolation problem consists of two subproblems: finding the
interpolant $\mathcal{P}_f$ and evaluating it on an assigned set of
points. The coefficients $c_j$ in (\ref{eq:base_exp}) are obtained
by imposing the interpolation conditions (\ref{eq:interp_cond})
$$\mathcal{P}_f(x_i)=\sum_{j=1}^N c_j\varphi(\Vert x_i-x_j\Vert)=y_i,\ i=1,\dots,N.$$
This leads to solve a linear system of equations (\ref{eq:RBFsys}).
\begin{figure*}[!t]
\begin{equation}
\label{eq:RBFsys}
\underbrace{\begin{bmatrix} \varphi(\Vert x_1-x_1\Vert) & \varphi(\Vert x_1-x_2\Vert) & \cdots & \varphi(\Vert x_1-x_N\Vert)\\ \varphi(\Vert x_2-x_1\Vert) & \varphi(\Vert x_2-x_2\Vert) & \cdots & \varphi(\Vert x_2-x_N\Vert)\\ \vdots & \vdots & \ddots & \vdots\\ \varphi(\Vert x_N-x_1\Vert) & \varphi(\Vert x_N-x_2\Vert) & \cdots & \varphi(\Vert x_N-x_N\Vert)\end{bmatrix}}_{A}  \cdot \underbrace{\left[ \begin{array}{c} c_1 \\ c_2 \\ \vdots \\ c_N \end{array} \right]}_{x} = \underbrace{\left[ \begin{array}{c} y_1 \\ y_2 \\ \vdots \\ y_N \end{array} \right]}_{b}
\end{equation}
\end{figure*}

Given a set of $M$ points $\mathbf{\xi}=\{\xi_1,\xi_2,\dots,\xi_M\}$
the evaluation of the interpolant $\mathcal{P}_f$ on $\xi$ can be
computed with a matrix vector product (\ref{matxvet})
\begin{figure*}[!t]
\begin{equation}\label{matxvet}
\left[ \begin{array}{c} \mathcal{P}_f(\xi_1) \\ \mathcal{P}_f(\xi_2) \\ \vdots \\ \mathcal{P}_f(\xi_M) \end{array} \right] =  \begin{bmatrix}\varphi(\Vert \xi_1-x_1\Vert) & \varphi(\Vert \xi_1-x_2\Vert) & \cdots & \varphi(\Vert \xi_1-x_N\Vert)\\ \varphi(\Vert \xi_2-x_1\Vert) & \varphi(\Vert \xi_2-x_2\Vert) & \cdots & \varphi(\Vert \xi_2-x_N\Vert)\\ \vdots & \vdots & \ddots & \vdots\\ \varphi(\Vert \xi_M-x_1\Vert) & \varphi(\Vert \xi_M-x_2\Vert) & \cdots & \varphi(\Vert \xi_M-x_N\Vert)\end{bmatrix}  \cdot \left[ \begin{array}{c} c_1 \\ c_2 \\ \vdots \\ c_N \end{array} \right]
\end{equation}
\end{figure*}

The RBF interpolant determination consists to solve a linear system
of equations $Ax=b$. In order to have a well-posed  problem the
matrix $A$ must be non-singular. Unfortunately no one has yet
succeeded in characterizing the class of all basic function
$\varphi$ that generates a non-singular matrix for arbitrary set
$\mathcal{X}=\{x_1,\dots,x_N\}$ of distinct data sites. The
situation is however much better with \emph{positive definite
matrices}. An important property of positive definite matrices is
that all their eigenvalues are positive, and therefore a positive
definite matrix is non-singular. Popular radial basis function
$\Phi$, that give rise to positive definite interpolation matrices,
are summarized in Table \ref{tab:RBFs}, we focused our work (as
\cite{yokota2010petrbf}) on the Gaussian function taking advantage
of its property as described in \S 3.
\begin{table*}
\label{tab:RBFs}
\caption{Examples of radial basis functions.}
\centering
\begin{tabular}{ c c c}
\hline
RBF & $\Phi$ &\ \\
\hline
Poisson radial function & $\dfrac{J_{s/2-1(\Vert x-x_j \Vert)}}{\Vert x-x_j\Vert^{s/2-1}}$ & $s\geq 2$\\
Inverse Multiquadric & $(1+\Vert x- x_j\Vert^2 )^{-\beta}$ & $\beta>\frac{s}{2}$\\
Mat\`ern function & $\dfrac{K_{\alpha}(\Vert x-x_j \Vert) \Vert x-x_j \Vert^{\alpha}}{2^{\beta-1}\Gamma(\beta)}$ & $\alpha=\beta-\frac{s}{2}>0$\\
Whittaker function & $\int_0^{+\infty}(1-\Vert x-x_j\Vert)_+^{k-1}t^\alpha e^{-\beta t}dt$ & $k=2,3,\ldots,\ \ \alpha=0,1,\dots$\\
Gaussian function & $\dfrac{1}{\sqrt{2\pi}\sigma}e^{\frac{-\Vert x-x_j\Vert^2}{2\sigma^2}}$ & $\sigma>0$\\
\hline
\end{tabular}
\end{table*}



\section{Parallel Surface Reconstruction}
A brief description of the overall surface reconstruction algorithm is listed below:

\begin{algorithm}
    \caption{Surface Reconstruction}
    \label{alg:surfrec}
\texttt{\textbf{Requirements:}}\\
\texttt{ point cloud $\mathcal{X}$, surface normals $\mathbf{n}_i$, evaluation grid $\xi$}
    \begin{algorithmic}[1]
        \STATE \texttt{ compute extended data set:\\ \ \ \ \
        $\mathcal{X}_{ext}=\mathcal{X}\bigcup\mathcal{X}_\delta^+\bigcup\mathcal{X}_\delta^-$
        by using $\mathbf{n}_i$;}
        \STATE \texttt{ find the interpolant $\mathcal{P}_f$ on $\mathcal{X}_{ext}$;}
        \STATE \texttt{ evaluate $\mathcal{P}_f$ on $\xi$;}
        \STATE \texttt{ render the surface;}
    \end{algorithmic}
\end{algorithm}

The steps 1 and 2 have been already discussed in \S 2.1; the step 3
requires a matrix vector multiplication as described in \S 2.2 and
the final step can be simply accomplished using the MATLAB software
with the command \texttt{isosurface} or other specific tool for the
rendering. The most computational expensive step is the second one,
which requires the solution of a system of $3N$ linear equation,
where $N$ is the initial point cloud size, as described in \S 2.2.
In the following section we describe the approach used to handle
this problem.

\subsection{Adopted solution}
\begin{figure}[h!]
\center
  \includegraphics[width=.45\textwidth]{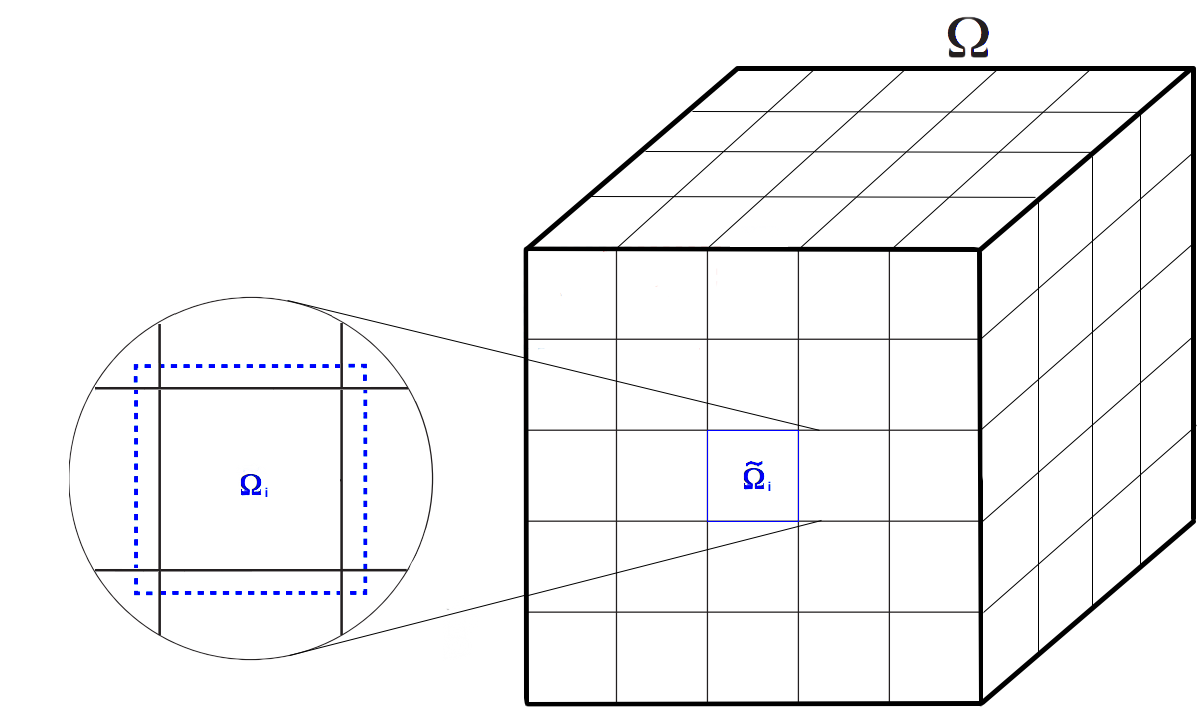}\\
  \caption{Illustration of the Domain Decompostion Method.}\label{img:ddm}
\end{figure}

Handling problems with large numbers of data points, as in our case
for surface reconstruction from clouds of millions of points, the
large amount of memory usage can become a problem. As the problem
size grows, parallelization on distributed memory architectures
becomes necessary. We adopt the idea behind Domain Decomposition
Methods (DDM) that is to divide the considered domain into a number
of subdomains and then try to solve the original problem as a series
of subproblems that interact through the interfaces. Let consider
the domain $\Omega$  containing the point cloud; the adopted domain
decomposition method divides the domain $\Omega$ in overlapping
sub-domains $\Omega_i$. The corresponding empty intersection
portions of subdomains are denoted with $\tilde{\Omega}_i$, as shown
in the example in Fig. \ref{img:ddm}.
The solution of the linear system $Ax = b$ on the whole domain can be obtained by sequentially solving, in the individual overlapping subdomains $\Omega_i$, the linear sub-system $A_ix_{\Omega_i} = b_{\Omega_i}$, where
$A_i$, $x_{\Omega_i}$ and $b_{\Omega_i}$ are the sub-elements corresponding to domain $\Omega_i$ for $A$,
$x$, and $b$ respectively. When each subdomain is solved individually and
the solution of the entire domain is updated simultaneously at the end of each iteration step, the method is called
\emph{additive} Schwarz method. Moreover, when the values $x$ outside of the subdomain $\tilde{\Omega}_i$ are discarded after the calculation of each subdomain $\Omega_i$, it is called \emph{restricted additive} Schwarz method (RASM). The RASM is
known to converge faster than the additive Schwarz method and requires less communication in parallel calculations. Furthermore solving smaller systems of equations has the same effect as a preconditioner, and then it can be used in combination with any iterative method like the Krylov subspace methods. In this work we use the Generalized Minimum Residual (GMRES).

Using basis functions with negligible global effects, the matrix $A$ can be considered
to have a finite bandwidth. In this case, the calculation of the matrix-vector multiplication, which is the predominant operation of the iterative solver, can be done somewhat locally. Using the Gaussian function as the basic function the matrix $A$ has the following elements:
\begin{equation}
\label{eq:matrixEntries}
A_{ij}=\frac{1}{\sqrt{2\pi}\sigma}\exp{\left(\frac{-\Vert \mathbf{x}_i-\mathbf{x}_j\Vert^2}{2\sigma^2}\right)}.
\end{equation}
In this way, since the Gaussian function decays rapidly, the elements of matrix
$A$ corresponding to the interaction of distant points can be neglected.
This sparsity of $A$ depends on the relative size of the calculation domain compared to the standard deviation, $\sigma$, of the Gaussian function. If $\sigma$ is kept constant while the size of the calculation domain is increased along with $N$, the calculation load will scale as $\mathcal{O}(N)$. The communication required to perform $Ax_i$ is also limited to a constant number of elements in the vicinity. Therefore, the RASM becomes an extremely parallel algorithm with minimum communication for the surface reconstruction problems that have a domain size of hundreds (or even thousands) of sigmas.

\subsection{Implementation details}
The algorithm implementation has been realized using the Portable,
Extensible Toolkit for Scientific Computation (PETSc)
\cite{balay1998petsc}. PETSc is a scalable solver library developed
at Argonne National Laboratory. All vectors and matrices can be
distributed by PETSc and each process stores only a local portion.
This transparent management of PETSc allows users to develop
scalable parallel code (almost) as serial code. In particular this
is done by using the \texttt{Vec} PETSc object for vectors $x$ and
$b$. To handle the overlapping and non-overlapping subdomain the
index sets (\texttt{IS}) were used. The PETSc \texttt{IS} object is
a global index that is used to define the elements in each subdomain
and is distributed among the processes in the same way as the
vectors. The interpolation matrix has entries which depends only
from the vector $\mathbf{x}$ and the Gaussian (eq. \ref
{eq:matrixEntries}). For this reason it is possible to use the
\texttt{MatShell} object which allow to make operations on the
matrix without actually storing the matrix. All calculation of inner
products, norms, and scalar multiplications are done by calling
PETSc routines and the linear system solution is finally calculated
with \texttt{KSPSolve}.

\subsection{GPU implementation}
GPU support has recently been added to PETSc to exploit the
performance of GPUs, these chips are highly optimized for
graphics-related operations. We use the CUDA framework that greatly
simplifies the programming model for GPUs. The GPU implementation of
PETSc also uses some of those libraries. Instead of writing
completely new CUDA code, PETSc uses the open source libraries CUSP
\cite{Cusp} e Thrust \cite{Thrust}. This allows transparent
utilization of the GPU without changing the existing source code of
PETSc.

A new GPU specific Vector and Matrix classes called \texttt{VecCUSP}
and \texttt{MatCUSP} has been implemented in PETSc. The classes use
CUBLAS, CUSP, as well as Thrust library routines to perform matrix
and vector operations on the GPU. The idea behind  these libraries
is to use already developed, fine tuned CUDA implementations with
PETSc instead of developing new ones. The PETSc implementation acts
as an interface between PETSc data structures and the external CUDA
libraries Thrust and CUSP.

Using the \texttt{VecSetType()} and \texttt{MatSetType()} PETSc routines, users can switch to the GPU version of the application simply using the command line parameters \texttt{-vec\_type cusp} and \texttt{-mat\_type aijcusp}.
Cusp natively supports several sparse matrix formats:
\begin{itemize}
\item Coordinate list (COO)
\item Compressed Sparse Row (CSR)
\item Diagonal (DIA)
\item ELLPACK (ELL)
\item Hybrid (HYB)
\end{itemize}
This feature is still in development in PETSc, however it can be used specifying \texttt{--download-txpetscgpu --with-txpetscgpu=1} in the configuration and compilation phase of PETSc.
After that it is possible for the application to switch between the sparse matrix formats with the command line option\\
\texttt{-mat\_cusp\_storage\_format <format>}.

A simple GPU implementation can be obtained by passing to the \texttt{MatShell} a pointer to a function that call a CUDA kernel that execute the operations on the Matrix.
A better idea is to use the GPU version of PETSc objects whenever possible, in order to accelerate all the available operation, and not only those on the matrix.
To this end there is the need of building the matrices object.
The Matrix for the interpolant determination (step 2) can be easily constructed using the algorithm reported in Algorithm \ref{alg:matrixConstruction}.
For the interpolant evaluation (step 3) instead the only required operation is the matrix vector multiplication. For this reason the construction of the matrix and the following matrix-vector multiplication
done directly by PETSc using CUSP is less efficient than a custom matrix-vector multiplication CUDA kernel (Algorithm \ref{alg:matrixVectorCUDA}) that consider the well known structure of the matrix (eq. \ref{eq:matrixEntries}) without the need of
building the Matrix.

\begin{algorithm}
    \caption{Interpolation matrix construction Pseudo-code algorithm }
    \label{alg:matrixConstruction}
    \begin{algorithmic}[1]
        \FOR{\textbf{each} subdomain $\Omega_i$}
            \FOR{\textbf{each} point $x_i$ in the  subdomain $\Omega_i$}
                \FOR{\textbf{each} point $x_j$ in the truncation area of $\Omega_i$}
                    \STATE Set $A_{ij}=\frac{1}{\sqrt{2\pi}\sigma}\exp{\left(\dfrac{-\Vert x_i-x_j\Vert^2}{2\sigma^2}\right)}$
                \ENDFOR
            \ENDFOR
        \ENDFOR
    \end{algorithmic}
\end{algorithm}

\begin{algorithm}
    \caption{CUDA code of RBF evaluation algorithm}
    \label{alg:matrixVectorCUDA}
    \begin{algorithmic}[1]
        \STATE \texttt{\textbf{\_\_shared\_\_ float} sharedXi[BLOCK\_SIZE];}
        \STATE \texttt{\textbf{\_\_shared\_\_ float} sharedGi[BLOCK\_SIZE];}
        \STATE \texttt{\textbf{int} bx = blockIdx.x;}
        \STATE \texttt{\textbf{int} tx = threadIdx.x;}
        \STATE \texttt{\textbf{int} i =  blockIdx.x * BLOCK\_SIZE +  threadIdx.x;}
        \STATE \texttt{\textbf{float} pf = 0;}
        \STATE \texttt{\textbf{float} coeff = 0.5f/(sigma*sigma);}
        \STATE \texttt{\textbf{for} (\textbf{unsigned int} m = 0; m < (col-1)/BLOCK\_SIZE+1; m++) \{}
                \STATE \texttt{sharedXi[tx] = Xi[m*BLOCK\_SIZE + tx];}
                \STATE \texttt{\_\_syncthreads();}
                \STATE \texttt{\textbf{for} (\textbf{unsigned int} k = 0; k < BLOCK\_SIZE; k++) \{}
                        \STATE \texttt{dx = Xj[i]-sharedXi[k];}
                \STATE \texttt{pf += sharedGi[k]*exp(-(dx*dx)*coeff;}\texttt{\}}\texttt{\}}
            \STATE \texttt{Pf[i] = pf/M\_PI*coef;}
    \end{algorithmic}
\end{algorithm}

\section{Experimental results}
In this section we present some results of our method for surface
reconstruction. The results presented here were computed using a
system equipped with an Intel Core i7-940 CPU (2,93 GHz, 8M Cache).
The middleware framework is OS Linux kernel 2.6.32-28 and PETSc
developer version 3.3.

Tests has been conducted to investigate the impact of the parameter $\sigma$ on the quality of the reconstruction.
As showed in  \S 3.2 our method is highly efficient for small value of $\sigma$ compared to the domain size.
However, besides efficiency, the quality of the result also depends strongly on the value of $\sigma$. This is because the accuracy of the interpolation model depends strongly on the ratio between the density of the point cloud and $\sigma$.

In \cite{yokota2010petrbf} experiments are carried out only on
equally-spaced lattice point distributions. In this particular case
they used as a measure for the density the spacing $h$ between the
points, discovering that a good choice for $\sigma$, in term of
performance and accuracy, is the one that satisfy the ratio
$h/\sigma\approx1$.

\begin{figure}
\center
  \includegraphics[height=.25\textheight]{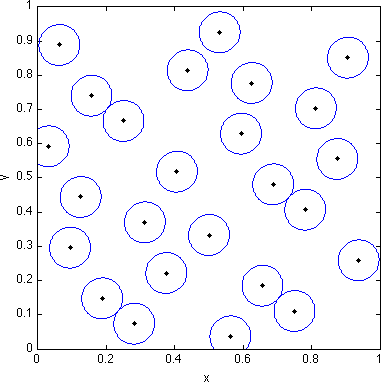}\hspace{.3cm}\includegraphics[height=.25\textheight]{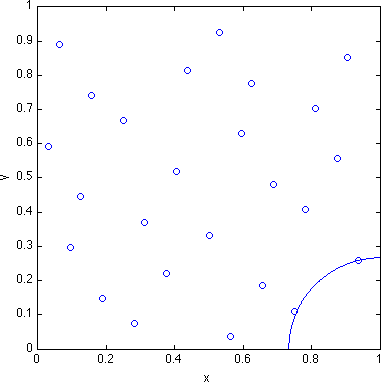}
  \caption{Geometric interpretation of separation distance (on the left) and fill distance (on the right) for 25 Halton points on the domain $\Omega=[0,1]^2$ ($q_\mathcal{X}\approx 0.0597$ and $h_{\mathcal{X},\Omega}\thickapprox 0.2667$).}\label{fig:filDist}
\end{figure}

For widely scattered data, as in the case of points cloud for surface reconstruction, there are more appropriate density measure available.
The first is the so-called \emph{separation distance} defined as
\begin{equation}
\label{eq:sepDist}
q_\mathcal{X}=\frac{1}{2}\min_{i\neq j}\Vert x_i -x_j\Vert_2.
\end{equation}
As shown in Fig. \ref{fig:filDist}, $q_\mathcal{X}$ geometrically represents the radius of the largest (hyper)sphere that can be drawn around each point in such a way that no (hyper)sphere intersects the others, which is why it is also sometimes called \emph {packing radius}. Another measure, usually used in approximation theory, is the so-called \emph{fill distance}:
\begin{equation}
\label{eq:fillDist}
h_{\mathcal{X},\Omega}=\sup_{x\in\Omega}\min_{x_j\in\mathcal{X}}\Vert x-x_j\Vert_2.
\end{equation}
It indicates how well the set $\mathcal{X}$ fill the domain $\Omega$. A geometric interpretation of the fill distance is given by the radius of the biggest empty (hyper)sphere that can be placed among the data locations in $\Omega$ (see Fig. \ref{fig:filDist}), for this reason it sometimes is also used as synonym the term \emph{covering radius}.
Hence for scattered data, using (\ref{eq:sepDist}) and
(\ref{eq:fillDist}), the heuristic ``optimal'' ratio
$h/\sigma\approx 1$ assumes the following expressions:
\begin{equation}
\label{eq:h1} \sigma\approx 2q_\mathcal{X}
\end{equation}
and
\begin{equation}
\label{eq:h2} \sigma\approx h_{\mathcal{X},\Omega}\sqrt{2}.
\end{equation}

 \subsection{Tests on synthetic dataset}
\begin{figure}[h!]
\center
  \includegraphics[height=.2\textheight]{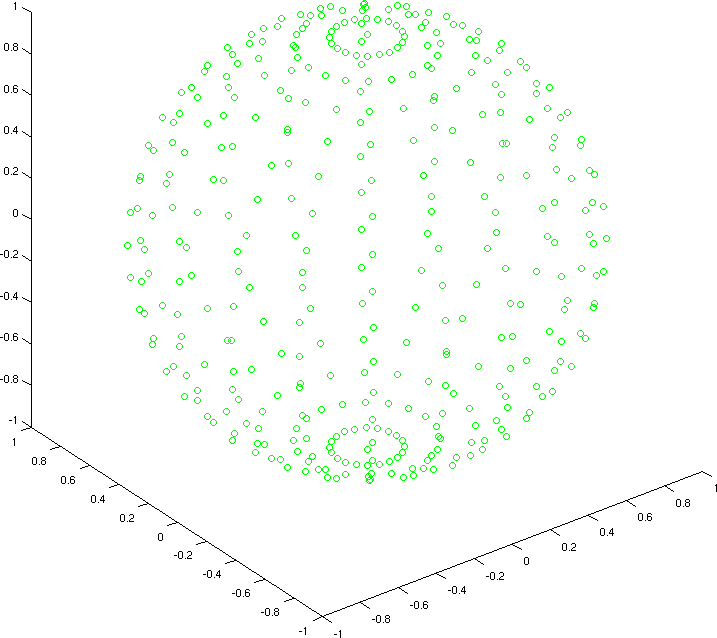}
  \caption{382 point cloud from the unit radius sphere centered in the origin.}\label{fig:sfera}
\end{figure}

We first conduct some experiment on a synthetic dataset. We choose
as surface a sphere, whose geometry is well know, and on which it is
possible to calculate (\ref{eq:sepDist}) and (\ref{eq:fillDist}). In
order to perform a consistent test with real dataset, we use a
widely scattered point cloud reported in Fig. \ref{fig:sfera}.

\begin{figure}
\center
\includegraphics[width=.22\textwidth]{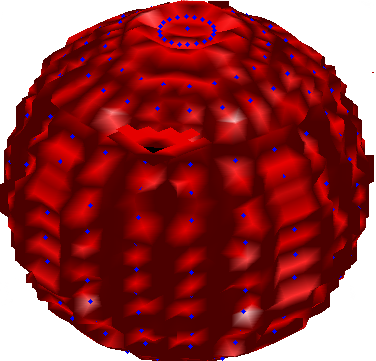}
\includegraphics[width=.22\textwidth]{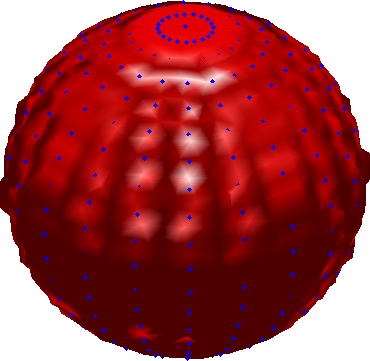}
\includegraphics[width=.22\textwidth]{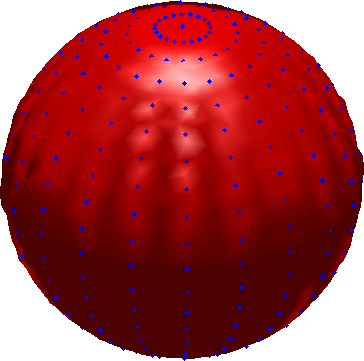}
\includegraphics[width=.22\textwidth]{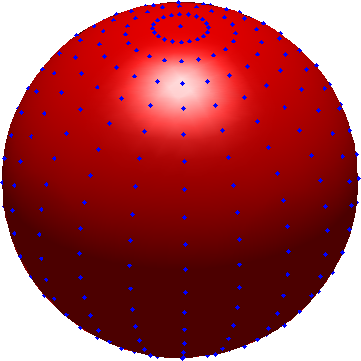}
\caption{Reconstructed sphere for different values of $\sigma$.
(from left to right $\sigma=0.025$, $\sigma=2q_\mathcal{X}=0.048$,
$\sigma=0.065$, $\sigma=h_{\mathcal{X},\mathcal{M}}\sqrt{2}=0.157$)}
\label{img:sfRecon}
\end{figure}

As shown in Fig. \ref{img:sfRecon} (top left) a value of $\sigma$
too small leads to a surface that actually interpolates the given
point cloud but whose reconstruction quality is total
unsatisfactory. Neither using (\ref{eq:h1}), as in
\cite{yokota2010petrbf}, is enough for scattered data (see Fig.
\ref{img:sfRecon}, top right). Increasing the value of $\sigma$
(Fig. \ref{img:sfRecon}, bottom left) the quality of the
reconstruction improves up to the desired one (Fig.
\ref{img:sfRecon}, bottom right) using (\ref{eq:h2}). It is
interesting to note that for intermediate values of $\sigma$, though
the quality is not globally satisfactory, it is locally sufficient
for areas where the points are at a distance $d \approx \sigma$.
\subsubsection{Tests on incomplete data}
\begin{figure}
\center
  \includegraphics[width=.25\textwidth]{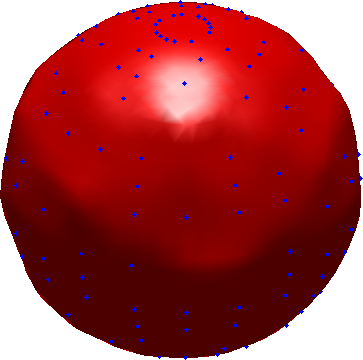}
\hspace{1cm}
  \includegraphics[width=.25\textwidth]{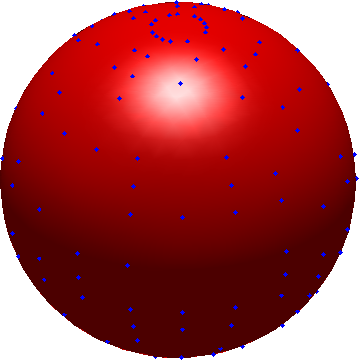}
\caption{Reconstructed sphere from incomplete data. (on the left
$\sigma=0.157$ and on the right
$\sigma=h_{\mathcal{X},\mathcal{M}}\sqrt{2}=0.328$)}
\label{img:miss}
\end{figure}

\ \\
We test the sensitivity of the algorithm to the lack of information
using incomplete dataset. We first use a dataset composed of $50\%$
randomly chosen points of the previous dataset. As shown in Fig.
\ref{img:miss} the point cloud is changed, then also the ``optimal''
value for $\sigma$ changes and becomes
$$\sigma=h_{\mathcal{X},\mathcal{M}}\sqrt{2}=0.328. $$

We remark that the previous value of $\sigma=0.157$  leads to a
reconstructed surface which is not the desired sphere (Fig.
\ref{img:miss}, top). Instead, by using the optimal value
$\sigma=0.328$, corresponding to the actual fill distance for the
new dataset, we  successfully reconstruct the sphere once again
(Fig. \ref{img:miss}, bottom).

\begin{figure}[h!]
\center
  \includegraphics[width=.22\textwidth]{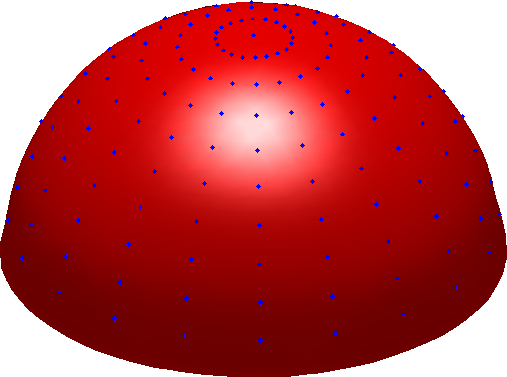}
\hspace{1cm}
  \includegraphics[width=.22\textwidth]{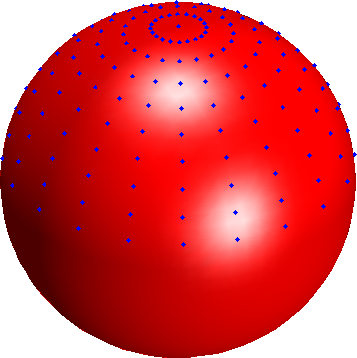}
\caption{Reconstructed surface from upper semi-sphere data. (on the
left $\sigma=h_{\mathcal{X},\mathcal{M}^+}\sqrt{2}=0.157$ and on the
right $\sigma=h_{\mathcal{X},\mathcal{M}}\sqrt{2}=2$)}
\label{img:half}
\end{figure}

For a second test we use a point cloud coming from the upper
semi-sphere:
$$
\mathcal{M^+}=\{(x,y,z)\in\mathbb{R}^3:\ x^2+y^2+z^2=1, z\geq 0\}.
$$
Using the value $\sigma=0.157$ which equals to
$h_{\mathcal{X},\mathcal{M}^+}\:\sqrt{2}$, as shown in
Fig.\ref{img:half} (top), we reconstruct with success the
$\mathcal{M^+}$ surface from which we took the data set. Instead, if
we assume that the point cloud came from the whole sphere
$\mathcal{M}$, the value of the fill distance changes to
$h_{\mathcal{X},\mathcal{M}}=\sqrt{2}$. Using this value  we get an
optimal value of $\sigma=h_{\mathcal{X},\mathcal{M}}\sqrt{2}= 2$,
which leads to the reconstruction of the whole sphere as shown in
Fig. \ref{img:half} (bottom).
\subsection{Tests on real dataset}
\begin{figure}[h!]
\center
  \includegraphics[width=.22\textwidth]{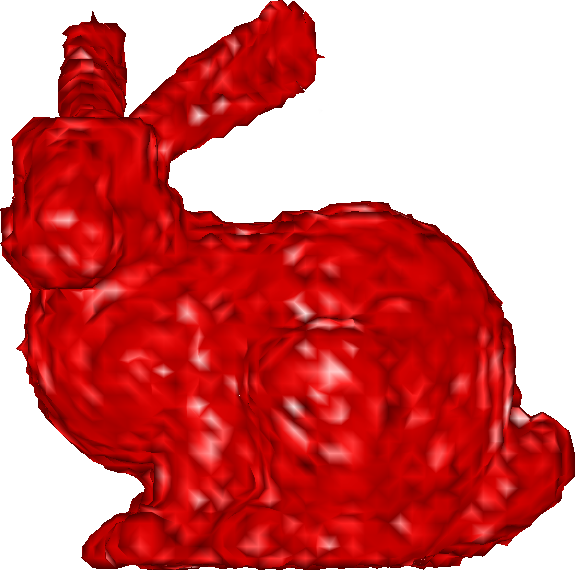}
  \includegraphics[width=.22\textwidth]{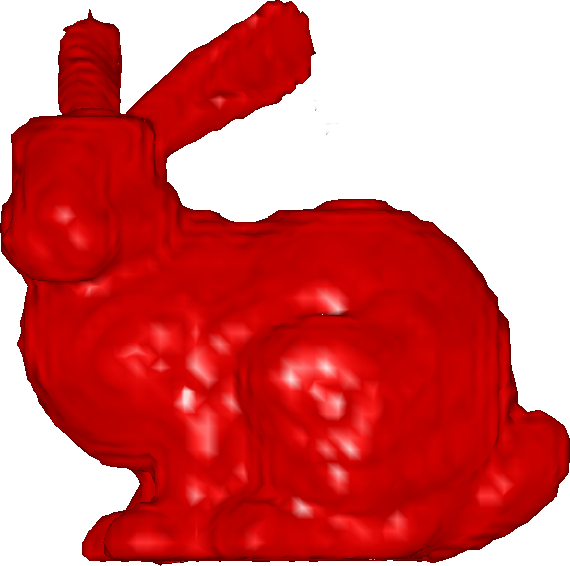}
  \includegraphics[width=.22\textwidth]{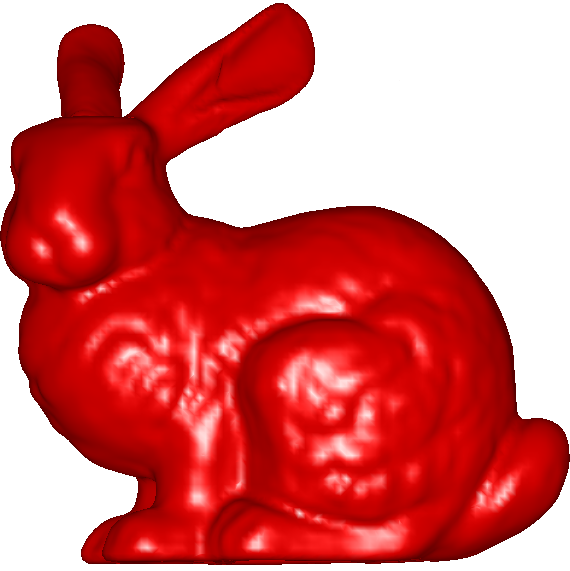}
\caption{Reconstructed bunny for different values of $\sigma$. (from
left to right $\sigma=0.0007$, $\sigma=0.0012$ and
$\sigma=h_{max}\approx h_{\mathcal{X},\mathcal{M}}\sqrt{2}=0.0033$)}
\label{img:bunny}
\end{figure}

Finally we test our optimal choice of $\sigma$ on real dataset using
the Stanford Bunny model. The dataset we used is composed of $N =
8171$ points which leads to an the extended dataset of $N_{ext} = 3N
= 24513$ points. In order to select the optimal value of $\sigma$
one need to calculate the value of the fill distance. On real
datasets, where the geometry of the surface in unknown or at least
really complex, this task is a real challenge. Recalling that the
fill distance measures the data density in the membership domain we
introduce a new measure defined as
$$
h_{max}=\max_j\min_{i\neq j}\Vert x_i -x_j\Vert_2.
$$
This represents the bigger distance among the distances calculated
between each point and its closer point. For a dataset without
multiple \emph{connected components}, the fill distance can be
approximated by
\begin{equation}
\label{eq:hmax}
h_{\mathcal{X},\mathcal{M}}\approx h_{max}\sqrt{2}.
\end{equation}
As in synthetic data set, a value of $\sigma$ too small results in a
low quality reconstructed surface (Fig. \ref{img:bunny}(top left)),
intermediate values leads to reconstructions with locally good
reconstruction for areas with high density points (Fig.
\ref{img:bunny}, top right) and a using (\ref{eq:hmax}) a choice of
$\sigma\approx h_{\mathcal{X},\mathcal{M}}$ is the optimal one (Fig.
\ref{img:bunny}, bottom).

\subsection{Tests on performance}
In this section, we will discuss the performance of our GPU
implementation on a single node compared to the CPU implementation.
CPU times refers to the execution on one core of i7-940 CPU and the
GPU times refers to execution on a nVIDIA Fermi C1060 GPU with 4Gb
of RAM. The middleware software consists of PETSc developer version
3.3 compiled with GPU support, CUDA release 4.2, CUSP version 0.3.0
and Thrust version 1.5.2.

For our tests we used a synthetic point cloud dataset with an
increasing number $N$ of points. Increasing the number of points of
the cloud corresponds to increasing the density of the domain. For
this reason, since we want to emphasize the benefits of the GPU, we
used a constant value of $\sigma=0.157$. Otherwise with a value of
$\sigma$ scaled with the density the problem would scale as
$\mathcal{O}(N)$ making less noticeable the contribution of the GPU.

In Tab. \ref{tab:coeff} we report the execution time on a single CPU
and GPU and the resulting speed up for the determination of the
interpolant (step 2 of the reconstruction algorithm) varying the
size $N$ of the data. To make a fair comparison we fixed the number
of iteration of the GMRES to 50. Results in Tab. \ref{tab:coeff}
shows that, even if the RASM preconditioner is not available in
CUSP, exploiting the GPU the execution time can be greatly
decreased.
\begin{table} [h!]
\caption{Speed up with GPU for interpolant determination.}
\label{tab:coeff}
\centering
\begin{tabular}{|c|c|c|c|}
\hline
$N$ & CPU & GPU & Speed up\\
\hline
1323    &0,67551     &0,1404 & 4,81\\
\hline
4686    &34,567 &5,6472 & 6,12\\
\hline
15625   &451,75 &71,57 & 6,31\\
\hline
24036 & 5398,4 & 815,09 & 6,63\\
\hline
\end{tabular}
\end{table}

In Tab. \ref{tab:eval} are reported the execution time on CPU and GPU with the resulting speed up for the
interpolant evaluation (step 3 of the surface reconstruction algorithm) varying the size $N$ of the point cloud and the size $M$ of the evaluation grid.
As expected these execution time are substantially lower than
those for the previous problem but in this case the GPU is fully exploited obtaining greater speed ups.

\begin{table} [h!]
\caption{Speed up with GPU for evaluation of interpolant.}
\label{tab:eval}
\begin{tabular}{|c||c|c|c|c|c|}
\hline
  $M$\textbackslash $N$ & \  & 1323 & 4686 & 15625 & 24036 \\
\hline
\hline
   \multirow{3}{*}{15625} & CPU &0,11571&0,20605&0,78172&0,97539\\
  & GPU &0,012831&0,033589&0,12167&0,13699\\
  & Speed up &\textbf{9,01}&\textbf{6,13}&\textbf{6,42}&\textbf{7,12}\\
\hline
\multirow{3}{*}{125000} & CPU &0,60406& 1,5657  &5,9558 &7,433\\
  & GPU &0,029907&0,09974&0,32034&0,43697\\
  & Speed up &\textbf{20,19}&   \textbf{15,69}& \textbf{18,59}&\textbf{17,01}\\
\hline
\multirow{3}{*}{421875} & CPU &1,9098&  5,2612  &19,946&    25,037\\
  & GPU &0,084603&0,2892467&0,89695&1,1272\\
  & Speed up & \textbf{22,57} & \textbf{18,18}&\textbf{22,23}&\textbf{22,211}\\
\hline
\multirow{3}{*}{1000000} & CPU &4,4389& 12,505& 47,594& 59,501\\
  & GPU &0,18456&0,60741&2,0629&2,4968\\
  & Speed up &\textbf{24,05}&\textbf{20,58}&\textbf{23,07}&\textbf{23,83}\\
\hline
\multirow{3}{*}{1953125} & CPU &8,8515&24,431&92,589&116,39\\
  & GPU &0,35668&1,0662&3,9195&4,8035\\
  & Speed up &\textbf{24,81}&   \textbf{22,91}& \textbf{23,62}& \textbf{24,23}\\
\hline
\multirow{3}{*}{3375000} & CPU &15,018  &42,166 &160,17 &199,45\\
  & GPU &0,59846    &1,7525 &6,4671 &7,9468\\
  & Speed up &\textbf{25,09}&\textbf{24,06}&\textbf{24,76}&\textbf{25,09}\\
\hline
\end{tabular}
\end{table}

\section{Conclusion}
We have implemented a parallel implicit method based on radial basis
functions for surface reconstruction. This implementation relies on
parallel scientific libraries and is supported for the GPU device.
Since the reconstruction quality and the performances are strongly
related to the gaussian RBF parameter $\sigma$, we propose an
optimal and heuristic estimate based on some density measures of the
point cloud. Finally, the obtained speed-up and running time confirm
that the RBF interpolant can be a very effective algorithm for such
problem.

\end{document}